\documentclass[aps,pre]{revtex4}
\usepackage{graphicx}
\begin{document}
\title{Clustering and ensembles inequivalence \\
in the $\phi^{4}$ and $\phi^{6}$ mean-field Hamiltonian models}

\author{Thierry Dauxois$^{1}$\email{Thierry.Dauxois@ens-lyon.fr},
Stefano Lepri$^{1,2}$\thanks{lepri@avanzi.de.unifi.it}, Stefano
Ruffo$^{1,2}$\thanks{ruffo@avanzi.de.unifi.it,phone:+39-055-4796344
  phone, Fax:+39-055-4796342}}\altaffiliation{Also at INFN Sezione di
  Firenze.}

\affiliation{$^1$ Laboratoire de Physique, UMR-CNRS 5672, ENS Lyon,
46 All{\'e}e d'Italie, 69364 Lyon c{\'e}dex 07, France\\
$^2$ Dipartimento di Energetica, Universit{\`a} di Firenze and 
INFM, via S. Marta, 3, 50139 Firenze, Italy}

\date{\today}

\begin{abstract}
We investigate a model of globally coupled conservative
oscillators. Two different algebraic potentials are considered
that display in the canonical ensemble either a second 
($\phi^{4}$) or both a second and a first order phase transition 
separated by tricritical points ($\phi^{6}$). The stability of highly
clustered states appearing in the low temperature/energy region
is studied both analytically and numerically for the $\phi^{4}$-model.
Moreover, long-lived out-of-equilibrium states appear close to the second
order phase transition when starting with ``water-bag" initial conditions, 
in analogy with what has been found for
the Hamiltonian Mean Field (HMF) model. The microcanonical
simulations of the $\phi^{6}$-model show strong hysteretic effects
and metastability near the first-order phase transition and a
narrow region of negative specific heat.
\end{abstract}
\maketitle

\section{Introduction}

The treatment of long-range interacting systems remains a challenging
issue in thermodynamics and statistical mechanics \cite{padmanabhan}.
Serious theoretical difficulties arise because internal energy,
entropy and other thermodynamic quantities are no longer additive, i.e.
a part of a system has not the same thermodynamic properties of the
whole. This originates unusual effects, like negative specific heat
and the inequivalence of statistical ensembles even in the limit of
infinite number of particles (See Ref.~\cite{springer} for a recent
review emphasizing different examples such as gravitation, plasmas,
fluid mechanics,\ldots).  Relevant physical examples displaying such
anomalies are known in Newtonian gravity but also in plasma physics
(although in the latter case the screening of attractive and repulsive
Coulomb interactions may mitigate them).

As usual in theoretical physics, the study of simple toy models
proves to be of major importance to attack more complex and
realistic systems. In particular, simple mean-field models with
infinite-range interactions turned out to be extremely useful. In
spite of the fact that they are constantly used in statistical
mechanics to describe cooperative phenomena, it is somehow
singular that violation of additivity has hardly been recognized
in the past. A
reason for that lies perhaps in the fact that the thermodynamic
limit is performed resorting to saddle-point techniques, which puts
the Hamiltonian in the explicitly decoupled form, thus hiding the
difficulties inherent in the long-range interaction.  Indeed,
ensemble inequivalence (for example between microcanonical and
canonical ensemble) has been observed, producing effects like
negative specific heats, which are the counterparts of the ones
known in the gravitational context~\cite{padmanabhan}.

The advantage of such models is that their canonical thermodynamics
can be exactly derived by performing the mean-field limit (the
infinite $N$ limit at fixed volume), which is a reasonable surrogate
of the thermodynamic limit (the infinite $N$ limit at fixed density).
Contrary to the usual belief, an exact microcanonical solution is also
feasible for such non trivial Hamiltonians, using large deviations
techniques~\cite{Largedev}, but the results will be presented
elsewhere~\cite{largedeviations}. Here, for what concerns the
microcanonical ensemble, we will mainly limit ourselves to show the result of
numerical simulations, which, because of the mean-field nature of the
interaction, require {\it only} ${\mathcal O}(N)$ codes (instead of
the usual ${\mathcal O}(N^2)$).  Moreover, further insight can be
gained from solving the one-dimensional collisionless
Boltzmann-Poisson equation for the single-particle distribution
function, which becomes exact in the $N \to \infty$ limit (at all finite
times)~\cite{Spohn}.

In the present paper, we investigate, both analytically and
numerically, two simple mean-field models which we denote as ``$\phi^4$"
and ``$\phi^6$", which display respectively second ($\phi^4$) and first and
second order phase transitions separated by tricritical points ($\phi^6$)
in the canonical ensemble. Of the former model, we investigate in
addition the dynamical formation of clustered states at low
temperatures and we study their destabilization. The presence of
quasi-stationary out-of-equilibrium states is moreover revealed close
to the second order phase transition, in analogy with what is found
for the Hamiltonian-Mean-Field (HMF) model (see \cite{hmfspringer} for
a recent review). Concerning the $\phi^6$-model, we study the phase
diagram in the canonical ensemble and we report numerical simulations
of hysteretic effects near first-order phase transitions. We point out
the existence of a narrow region of negative specific heat.

\section{The mean-field $\phi^4$ model}

Let us first consider the following Hamiltonian
\begin{equation}
H = \sum_{i=1}^N \left[{p_i^2 \over 2} - (1-\theta) {q_i^2 \over 2}+
{q_i^4 \over 4}\right] - {\theta \over 2N}\sum_{i,j=1}^N  q_iq_j\quad,
\label{hamiltonian}
\end{equation}
where $p_i$ is the conjugate momentum of the variable $q_i$, which
defines the position of the $i$-th particle on a line.  This is a mean field
model since all particles are connected to all others, and the
summation in the last term is {\em not} restricted to neighboring
particles. Notice that positive (resp. negative) values of the
parameter $\theta$ correspond to attractive (resp. repulsive)
mean-field interactions. All variables are adimensional and, for
the sake of comparison, we have used the same parametrization
introduced in Ref.~\cite{zwanzig} (which can be shown to be
minimal by conveniently rescaling the variables and time). The
local potential displays a double well for $\theta<1$ and a single
well otherwise.  The ground-state energy per particle is
$e_0=-1/4$ for positive $\theta$ (all particles in a single
cluster) and $e_0=-1/4 +\theta/2$ in the repulsive case (double
cluster).

\subsection{Dynamics of the Magnetization: The generation of clusters.}

Introducing the time-dependent magnetization
\begin{equation}
\displaystyle
M=\frac{1}{N}\sum_{i=1}^{N}q_i\quad,
\end{equation}
we are therefore interested in the following equations of motion
\begin{equation}
\ddot q_i=(1-\theta)q_i-q_i^3+\theta M\quad .
\label{equationmotion}
\end{equation}

We study the dynamics of particle released with a water
bag~\cite{waterbag} initial condition where positions and momenta
are uniformly distributed at random in the intervals $[q_0-w_q/2,
q_0+w_q/2]$ and $[-w_p/2, +w_p/2]$, respectively.  We have adopted
the symplectic 6th-order Yoshida's algorithm~\cite{yoshida}, with
a time step $dt =0.05$, which allows us to obtain an energy
conservation with a relative accuracy $\Delta E / E$ ranging from
$10^{-7}$ to $10^{-10}$.

\begin{figure}
\null\hskip-16truecm
\resizebox{0.6\textwidth}{!}{\includegraphics{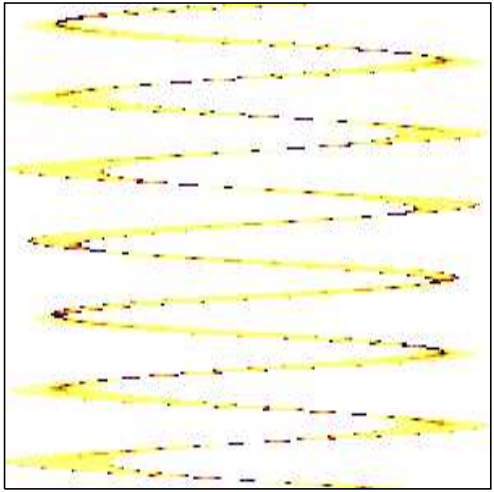}}
\caption{Dynamics of the cluster. Evolution of the density $\rho
(q)$
  of formula (\ref{density}) in grey scale for short times.  The darker the grey, the
  bigger the density. Space is horizontal, whereas the vertical downward
  direction corresponds to time. One notices the periodic motion
  with the characteristic time scale $\omega_M^{-1}$,
  defined in the text. In this simulation $N=4096$, $\theta=0.5$.}
\label{cluster}
\end{figure}

\begin{figure}[ht]
\begin{center}
\begin{tabular}{cc}
(a) & (b)\\
\includegraphics[width= 50mm,angle=-90]{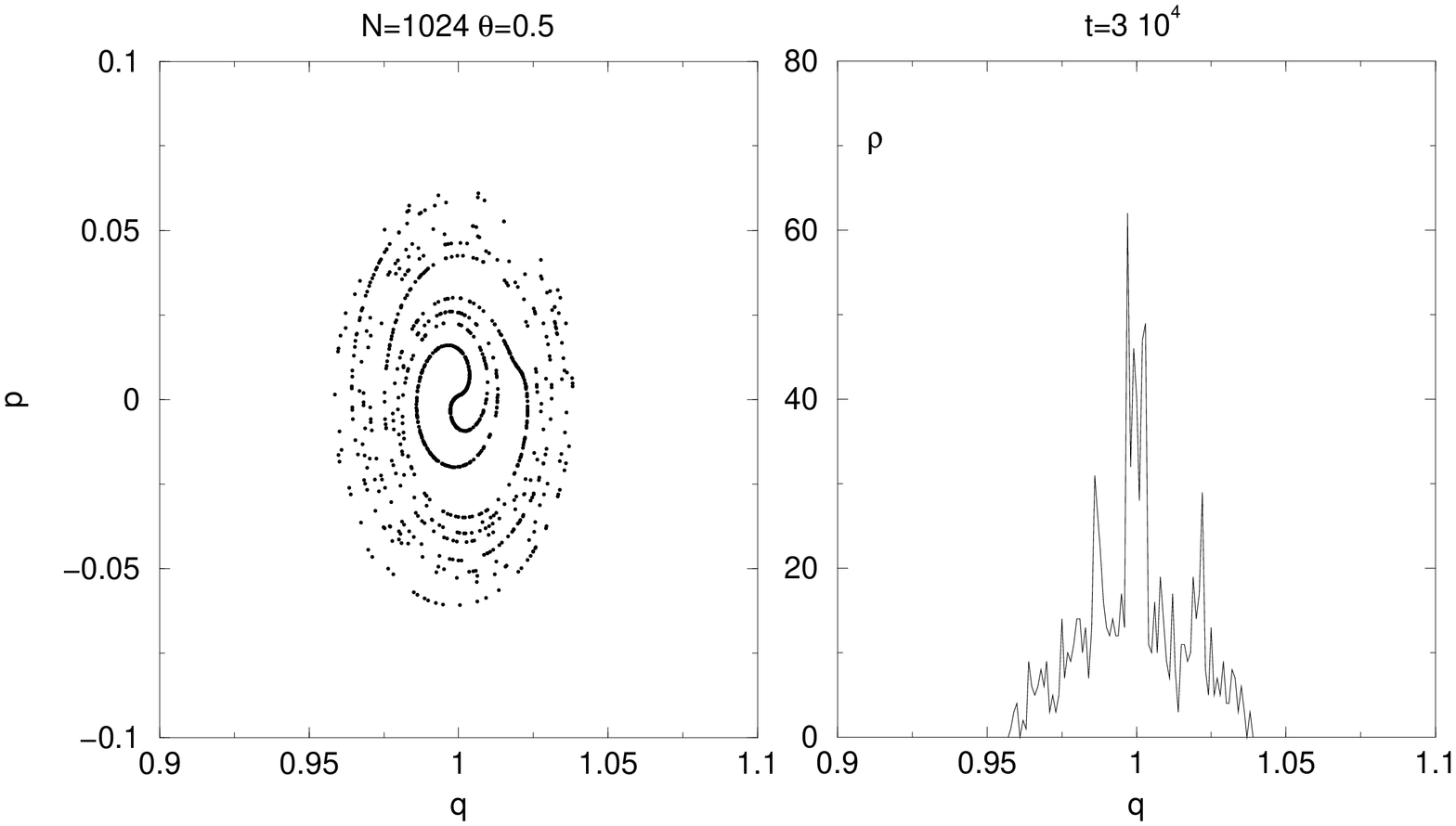} &
\includegraphics[width= 50mm,angle=-90]{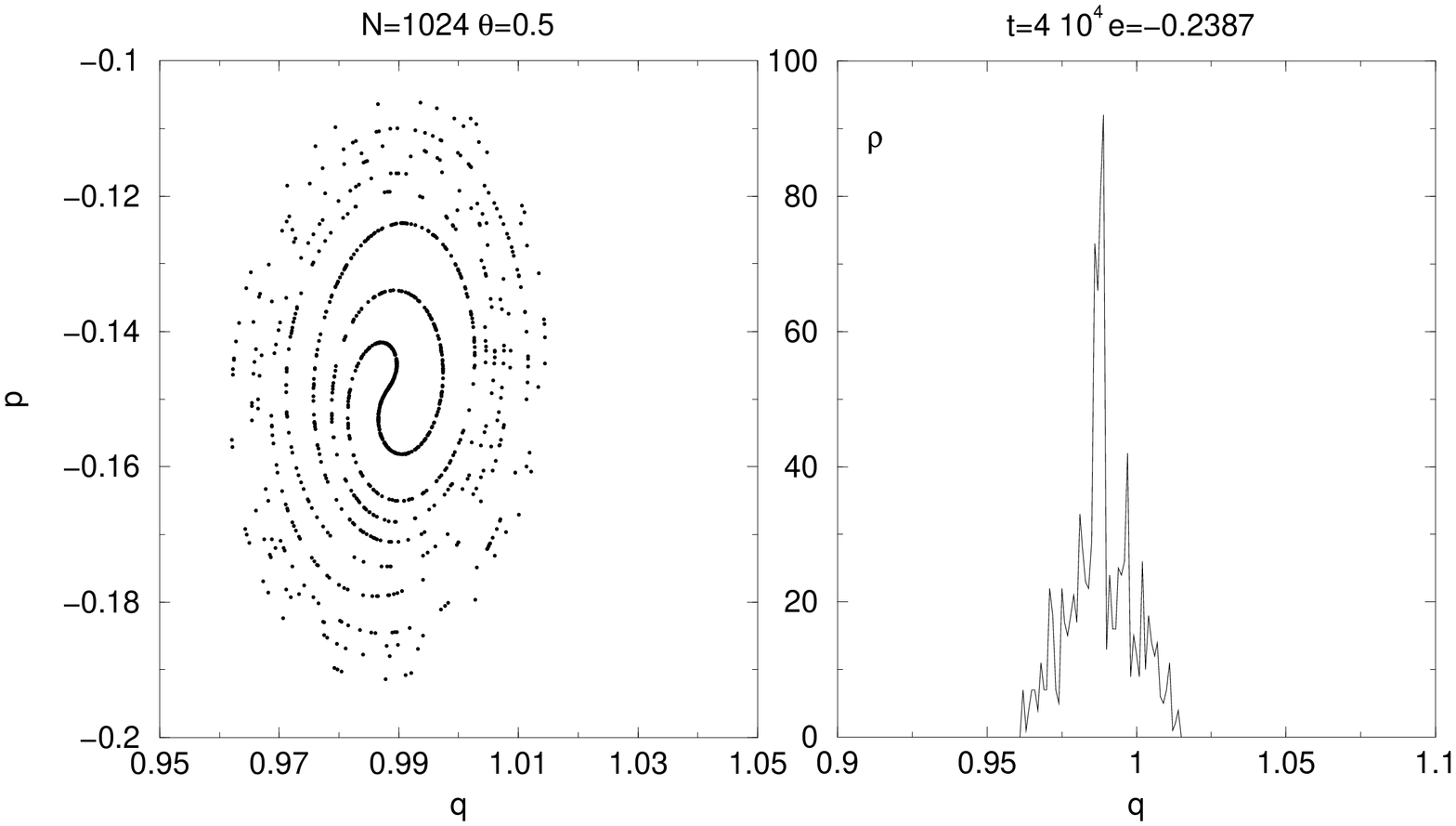} \\
\end{tabular}
\end{center}
\noindent\caption{Phase space snapshots of the cluster and the corresponding density
$\rho (q)$ of formula (\ref{density}) at two different times. In this simulation
$N=1024$, $e=-0.2387$,$\theta=0.5$}
\label{cluster2}
\end{figure}

Fig.~\ref{cluster} shows the result: a coherent oscillating
cluster self-consistently moving in the self-generated potential.
The data are obtained for an initial condition with a
small velocity dispersion, i.e. $q_0=1.1$, $w_q = 0.05$,
$w_p=0.0001$. Besides the oscillation of the center, the particles
display a rotating motion around it, which creates a spiral
structure (see Figs.~\ref{cluster2}), as frequently encountered in
long range systems; we have found this coherent behavior for a very
large collection of initial states. Notice that the spiral
structure in the center is responsible for the very large peaks in
the single particle density (right panels in
Figs.~\ref{cluster2}). A similar phenomenon has been described
successfully for the antiferromagnetic HMF model in terms of shock
waves \cite{barre,barreepjb} by considering the associated Vlasov
equation valid in the $N\to \infty$ limit.

\bigskip

We will therefore rely on a Vlasov-like approach.  Denoting by
$f(q,p,t)$ the one particle distribution function, we have here
\begin{equation}
\frac{\partial f}{\partial t}+p \frac{\partial f}{\partial q}
+\frac{\partial f}{\partial p}\left[(1-\theta)\ q-q^3+
\int_{-\infty}^{+\infty}du \int_{-\infty}^{+\infty} d\alpha \ f(\alpha,u,t)
\ \alpha\right] = 0\label{vlasov}\quad .
\end{equation}
Introducing a density field $\rho$ and a velocity field $v$, as follows
\begin{eqnarray}
\rho(q,t)& = & \int_{-\infty}^{+\infty}f(q,p,t)dp\label{density}\\
\rho(q,t)v(q,t)& = & \int_{-\infty}^{+\infty}pf(q,p,t)dp
\end{eqnarray}
and neglecting velocity dispersion, we have recently
shown~\cite{barre} how to reduce this problem to appropriate
hydrodynamical equations. A short-time analysis, performed for the
HMF model, led us finally to a {\em dissipativeless} spatially
forced Burgers equation. We expect that a similar treatment can be
developed for the current model and that similar techniques could be applied.
A well known property of the Burgers equation without viscosity, 
is that the solution becomes
multi-stream after a {\em finite} time: the appearance of shock
waves in the velocity profile corresponds indeed to singular
points in the density profile (see Figs.~\ref{cluster2}). In the
original {\em discrete} model, this phenomenon would correspond to
particle crossing; after some time, fast particles will eventually
catch slow ones downstream creating the so-called
spiral dynamics exemplified in the left panels of Figs.~(\ref{cluster2}).

\subsection{Stability analysis}

\begin{figure}
\centerline{\resizebox{0.4\textwidth}{!}{\includegraphics{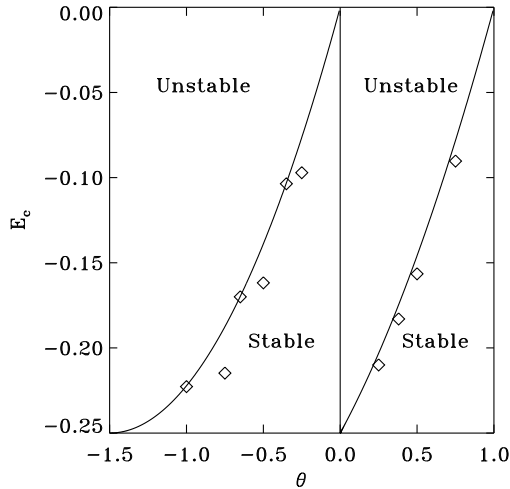}}}
\caption{Critical energy as a function of the parameter $\theta$.
  The solid line corresponds to the results given by the stability
  charts derived analytically (or, alternatively, using the Floquet
  analysis), whereas the diamonds correspond to the results of microcanonical
  simulations for an ensemble of $N=1024$ particles.}
\label{compatheorynumerique}
\end{figure}

To understand the origin of this cluster and its stability ,
we will first consider the simplest case of a fully clustered state
where $q_i=M$.  In this simple case, the collective motion is ruled by
the equation
\begin{eqnarray}
\ddot M&=&M-M^3~,
\label{equationmotionbb}
\end{eqnarray}
which can be easily solved using elliptic functions~\cite{lawden}.
Integrating Eq.~(\ref{equationmotionbb}), between the initial time, 
when the cluster is released
without kinetic energy at the position $q=a>1$, and time $t$,
we get
\begin{eqnarray}
M&=&a\ \mbox{dn}\left(\frac{at}{\sqrt{2}},k\right)~,
\label{resulforM}
\end{eqnarray}
where $\mbox{dn}$ is the elliptic delta amplitude function and
$k=\sqrt{2-2/a^2}$ the modulus of Jacobi elliptic functions. We remind
that this solution is periodic, with the amplitude-dependent period
given in terms of the complete elliptic integral of the first kind
$2K(k)$; the magnetization $M$ will thus oscillate with a frequency
$\omega_M={\pi a}/{\sqrt{2}K}$, which will be the main timescale of
the problem. One notices immediately that the modulus $k$ and the
frequency $\omega_M$ are both functions of the same parameter, namely the
amplitude $a$, related to the energy per particle $e=E/N$ through the
relation $a^2={1+\sqrt{1+4e}}$.
This solution is interesting in its own, since explicit analytical
solutions are not common for nonlinear non-integrable systems of
oscillators, but one should of course study its stability in order to
understand why this coherent oscillating cluster emerges spontaneously.
Using the equations of motion~(\ref{equationmotion}) for
the $q_i$ and introducing $\xi_i=q_i-M$, we obtain up to first order
\begin{eqnarray}
\ddot\xi_i+(\theta-1+3M^2)\ \xi_i&=&0\label{tangentmap}\quad .
\end{eqnarray}
Introducing the new variable $u={at}/{\sqrt{2}}$, we obtain the Lam{\'e}
equation in its canonical form
\begin{equation}
\frac{d^2 \xi_i}{d u^2}+\left[\alpha-\nu(\nu+1)k^2\
\mbox{sn}^2\left(u,k\right)\right]\ \xi_i=0~,
\label{peccelli}
\end{equation}
with $\alpha=6+{2}(\theta-1)/{a^2}$ and $\nu=2$. For integer values of
$\nu$, many rigorous results are known~\cite{whittaker,Arscott} and in
particular it is established that there are only $\nu+1$ instability regions
in the $(\alpha,k)$ plane.  The stability charts could be  explicitly
constructed~\cite{pecelli} by observing that Eq.~(\ref{peccelli})
has the following five periodic solutions
\begin{eqnarray}
y&=1-\frac{\alpha}{2}\mbox{sn} ^2(u,k)& \quad\mbox{with}\quad \alpha=2\left[1+k^2\pm(k^4-k^2+1)^{1/2}\right]\\
y&=\mbox{cn}(u,k)\mbox{dn}(u,k)& \quad\mbox{with}\quad \alpha=1+k^2\\
y&=\mbox{sn}(u,k)\mbox{dn}(u,k)& \quad\mbox{with}\quad \alpha=1+4k^2\\
y&=\mbox{sn}(u,k)\mbox{cn}(u,k)& \quad\mbox{with}\quad \alpha=4+k^2\quad.
\end{eqnarray}
Thus the above curves $\alpha=\alpha(k^2)$ define the boundary
curves of the three ($\nu+1$) non-degenerate instability regions.
Theses curves are presented in the plane $(\theta,e)$ in
Fig.~(\ref{compatheorynumerique}).

One can also investigate the linear stability of this cluster solution
with a standard Floquet analysis, i.e. computing the eigenvalues of
the $2N\times 2N$ matrix of the tangent map.  Here, contrary to usual
lattice systems with coupling between neighbors,
Eq.~(\ref{tangentmap}) shows that we obtain $N$ identical second order
equations; this is a direct consequence of the mean field character of
Hamiltonian~(\ref{hamiltonian}). Consequently, we obtain two
different $N$ times degenerate Floquet eigenvalues and the periodic
solution is linearly stable when the eigenvalues lie on the unit
circle in the complex plane.

At this stage, one derives numerically the
linear stability threshold by considering the numerical evolution of
two different initial conditions (1,0) and (0,1) for the vector
$(\xi,\dot \xi)$. The dynamics is solved by a standard 4th order
Runge-Kutta algorithm for the time integration of
Eq.~(\ref{tangentmap}), where the magnetization $M$ is either directly
integrated using Eq.~(\ref{equationmotionbb}) or implemented with the
help of Eq.~(\ref{resulforM}).  For a given value of $\theta$,
an energy threshold exists, above which the largest Floquet
multiplier is greater than unity, and therefore the solution is
unstable.  The solid line in Fig.~\ref{compatheorynumerique} shows
the evolution of this threshold as a function of the
parameter~$\theta$.
The analytical calculations were directly compared with the numerical
thresholds obtained by considering a water bag with very small but
finite width, i.e. $w_q\ll q_0$ and $w_p\ll 1$, to make a direct
comparison with the above analytical results. Checking on
Fig.~\ref{compatheorynumerique}, one gets, apart from a slight
underestimate, a good agreement between numerics and theory.
It should however be remarked that, for the finite $N$ systems,
stability persists only for a finite time, which presumably diverges as
$N$ increases, as it happens for the HMF model~\cite{hmfspringer,barreepjb}.

\subsection{Equilibrium Statistical Mechanics}

The partition function can be computed by means of a standard
Hubbard-Stratonovich transformation. Indeed, for a Hamiltonian of the
general form
\begin{equation}
H=\sum_{i=1}^N\left[\frac{p_i}{2}^2+ V(q_i)\right]-
\frac{\theta}{2N}\left(\sum_{i=1}^{N}q_i\right)^2\quad ,
\end{equation}
the partition function is
\begin{equation}
Z=\int_{-\infty}^{+\infty} \prod_{\ell=1}^N dp_\ell\
 dq_\ell\  e^{-\beta H}=Z_K Z_V
=\left(2\pi /\beta\right)^{N/2} Z_V~,
\end{equation}
where the configurational partition function is
\begin{equation}
Z_V \; = \;
\int_{-\infty}^{+\infty}\prod_{\ell=1}^N dq_\ell\
e^{-\displaystyle \beta V(q_\ell)}\
e^{\displaystyle  \frac{\beta\theta}{2N}\left(\sum_{i=1}^{N}q_i
\right)^2}\quad .
\end{equation}
We use at this point the Hubbard-Stratonovich trick, i.e. we consider
the identity
\begin{equation}
e^{\displaystyle\mu{x^2}}=\frac{1}{\sqrt{\pi}}\int_{-\infty}^{+\infty}dy\
e^{\displaystyle -y^2+2\sqrt{\mu}xy} \quad.
\end{equation}
Defining
\begin{equation}
\psi(x,\beta) \;=\;   \ln\left[
\int_{-\infty}^{+\infty} dq\  e^{-\displaystyle \beta V(q)+xq}\right]\quad,
\label{definitionofpsi}
\end{equation}after some algebra one gets
\begin{eqnarray}
Z_V&=&\sqrt{\frac{N}{2\beta\theta\pi}}\
\int_{-\infty}^{+\infty}dx\ e^{\displaystyle
-N\beta f_L(x,\beta)}\\
\beta f_L &=& \frac{x^2}{2\beta\theta}-\psi(x,\beta)
\label{landau}
\end{eqnarray}
where $f_L$ is the configurational Landau free energy. In the
thermodynamic limit, one can evaluate the above integral by means of the
saddle point approximation. The saddle point is determined by the
condition $ {\overline x}={\beta\theta}\ \psi_x({\overline x,\beta})$,
(where $\psi_{x}$ denotes the derivative with respect to $x$)
and can be evaluated numerically in a self-consistent manner. Notice that
 ~${\overline x}=0$ is always a solution if the potential $V$ is even.

Finally, we can thus express the configurational partition function as
\begin{equation}
Z_V
=\frac{1}{\sqrt{\beta\theta\psi_{xx}({\overline x},\beta) -1}}
\quad
\exp{\left[\displaystyle N\left(\psi({\overline x},\beta) -\frac{{\overline
x}^2}{2\beta\theta}\right)\right]}\quad .
\end{equation}
Up to terms of order ${\cal O}\left({1}/{N}\right)$,
the relevant equilibrium observables can be expressed accordingly as a
function of $ {\overline x}$, using the following formul{\ae}:
\begin{eqnarray}
\beta f &=&-\frac{1}{N}\ln Z=
-\frac{1}{2}\ln \left(2\pi /\beta\right)
- \psi({\overline x},\beta) +\frac{{\overline x}^2}{2\beta\theta} \\
M &=& \left\langle\frac{1}{N}\sum_{i=1}^{N}q_i  \right\rangle =
\frac{\overline x }{\beta\theta}\label{canoform} \\
e&=&\frac{\partial \beta f}{\partial\beta}=
\frac{1}{2\beta}- \psi_\beta ({\overline x},\beta)-
\frac{{\overline x}^2}{2\beta^2\theta}
=\frac{1}{2\beta}- \psi_\beta ({\overline x},\beta)-\frac{\theta}{2} M^2\quad .\label{canofore}
\end{eqnarray}
In the disordered phase, when $M=0$, the system reduces to an ensemble of
independent anharmonic oscillators.

\begin{figure}[ht]
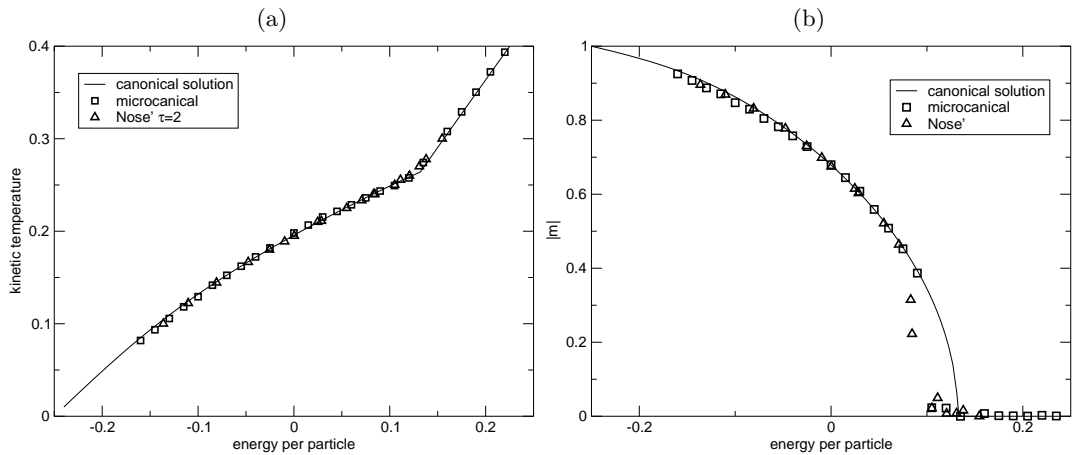

\begin{center}
\begin{tabular}{cc}
(a) & (b)\\
\includegraphics[clip,width= 70mm]{compare-ens.eps} &
\includegraphics[clip,width= 70mm]{compare-ens2.eps} \\
\end{tabular}
\end{center}
\noindent\caption{Comparison of ensembles for the $\phi^4$ model,
$\theta=0.5$, $N=512$:  caloric (panel a) curve and magnetization
(panel b). Squares and triangles refer to microcanonical and
canonical simulations, respectively, while the solid lines are the
exact canonical solutions given by Eqs.~(\ref{canofore})
and~(\ref{canoform}). The critical point is located at $T_c=0.264$
($e_C=0.132$). In both cases the initial conditions were
$q_i(0)=1$ and $p_i(0)$ chosen randomly with a gaussian
distribution.} \label{p4comp}
\end{figure}

In the ferromagnetic case ($\theta>0$), as presented in
Fig.~\ref{p4comp}, the model displays a second-order transition in
the canonical ensemble, in full agreement with previous results
based on the Fokker-Planck approach \cite{zwanzig}. The
magnetization vanishes as $(T_c-T)^{\frac{1}{2}}$ in the
subcritical regime and the specific heat has a finite jump at
$T_c$. Conversely, in the antiferromagnetic case ($\theta<0$), no
transition occurs, and $\overline x=0$ is the only solution of the
consistency equation for any value of the temperature. The free
energy and the internal energy are always given by the above
formul{\ae}, with zero magnetization.

As this behavior is clearly reminiscent of the HMF model, that we have already
studied in the past~\cite{antoni,dauxois,barre}, and where
the presence of long-lived out-of-equilibrium states
was surprisingly discovered, it was natural to suspect that they also
appear in the present model.  We have
therefore performed two types of numerical simulations: microcanonical
ones with a symplectic algorithm (sixth order Yoshida or fourth-order
McLachlan-Atela~\cite{macla}) and canonical ones with a
Nos{\'e}-Hoover thermostat~\cite{Nose} (fourth-order Runge-Kutta algorithm).
No appreciable deviations are observed between the two types of simulations
for initial conditions close to equilibrium, see Fig.~\ref{p4comp}.

\begin{figure}[ht]
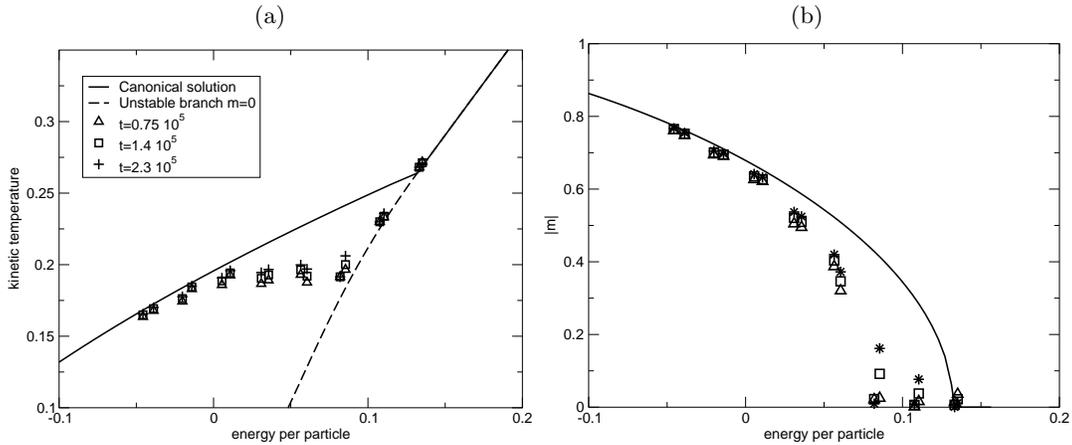

\begin{center}
\begin{tabular}{cc}
(a) & (b)\\
\includegraphics[clip,width=7cm]{wb-temp-N10000.eps}&
\includegraphics[clip,width=7cm]{wb-mag-N10000.eps}\\
\end{tabular}
\end{center}
\caption{Microcanonical results obtained using
molecular dynamics simulations  for the ferromagnetic $\phi^4$ model:
$\theta=0.5$, $N=10^4$ and water bag initial conditions $q_0=1$,
$w_q=10^{-4}$.}
\label{wbag}
\end{figure}

However, a region with clear differences is found for ``water-bag"
initial conditions: see an example for $q_0=0$ in Fig.~\ref{wbag}.
This is strongly reminiscent of similar observations made on the HMF
model~\cite{antoni,Tsallis}.
The fact that some points lie on the branch with vanishing
magnetization also in the subcritical region (see Fig.~\ref{wbag}b)
indicates that this is a metastable state in the microcanonical
ensemble. On the contrary, let us notice that triangles below the
theoretical curve in Fig.~\ref{p4comp}b are due to finite size effects
and would disappear for larger $N$ values.  A careful study of the
numerical results for very large integration times shows a
systematic tendency of these points to converge towards the
equilibrium state indicated in Fig.~\ref{p4comp} by the solid line. 
This attests the metastable character of these states.

Series of microcanonical runs for the repulsive case have shown that
metastable states may possibly exist also in this case, and we suspect
that they may be related to the existence of a stable cluster in the
energy region $-1/4< e \lesssim -0.097$.  This metastability is thus of
dynamical rather than of thermodynamical origin.


Summarizing, the $\phi^4$ model has emphasized the striking appearence of
a cluster and also dynamical differences between microcanonical and
canonical ensembles, presumably related to slow relaxation towards
the final Boltzmann-Gibbs equilibrium state; ensemble inequivalence
appears only in a transient. Indeed, it has been recently
reported in spin systems~\cite{beg}, that true ensemble inequivalence occurs in
regions of first order phase transitions. It would be therefore very
interesting to exhibit a {\em dynamical} mean field model of the
polynomial class with a first order phase transition. This is the purpose
of the next section.

\section{The mean-field  $\phi^6$ model}

The simplest generalization of the previous model is
\begin{equation}
H = \sum_{i=1}^N \left[{p_i^2 \over 2} +
r {q_i^2 \over 2}-{q_i^4 \over 4}+{q_i^6 \over 6} \right]-
{D \over 2N} \sum_{i,j=1}^N  q_iq_j\quad,
\label{modellofi6}
\end{equation}
where $D$ and $r$ are two independent parameters (also in this case
it can be shown that this parametrization is minimal).  The main interest 
of model (\ref{modellofi6}) lies in
the fact that it may exhibit a first-order phase transition for a
proper choice of the parameters, and therefore possibly ensemble
inequivalence.  This can be realized by first considering the zero temperature
limit, where equilibrium states are given by the minima of the function
\begin{equation}
V_{\mbox{eff}}={r-D \over 2} x^2 - {x^4 \over 4}+{x^6 \over 6}\quad .
\end{equation}
For $0< r-D < 1/4$, such polynomial admits three minima located at
$x=0$ and $x=\pm x_+$ and two maxima at $x=\pm x_-$ where
\begin{equation}
x_\pm^2 = {1\over 2} \pm {1\over 2} \sqrt{1-4(r-D)}\quad .
\end{equation}
A first-order transition can thus be expected within this parameter
region.  Furthermore, in order to have a first-order phase transition
at $T=0$ we must impose that the two minima attain the same value
(equal to zero). This conditions holds for $r-D=3/16$ and we can at
least hope that close to this parameter values the transition persists
also at nonzero temperature. We checked that this is indeed
the case by computing the free energy in a self-consistent way, as
explained in the previous section.
The transition exists in a very narrow region {\it below } $r-D=3/16$.

\begin{figure}[ht]
\begin{center}
\begin{tabular}{cc}
(a) & (b)\\
&\includegraphics[width=4cm,angle=-90]{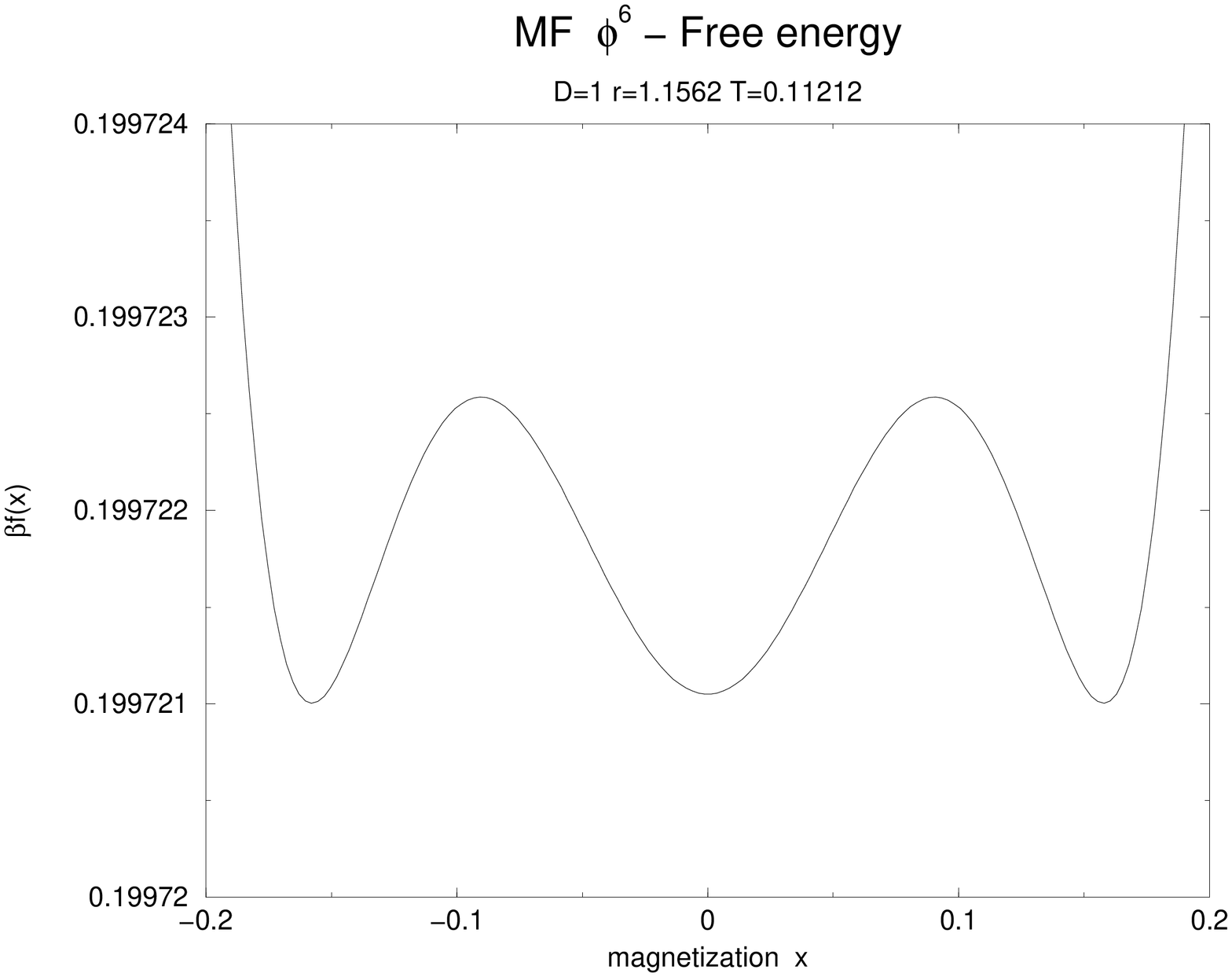}\\
\raisebox{3.cm}[0pt]{
\includegraphics[width=7cm,angle=-90]{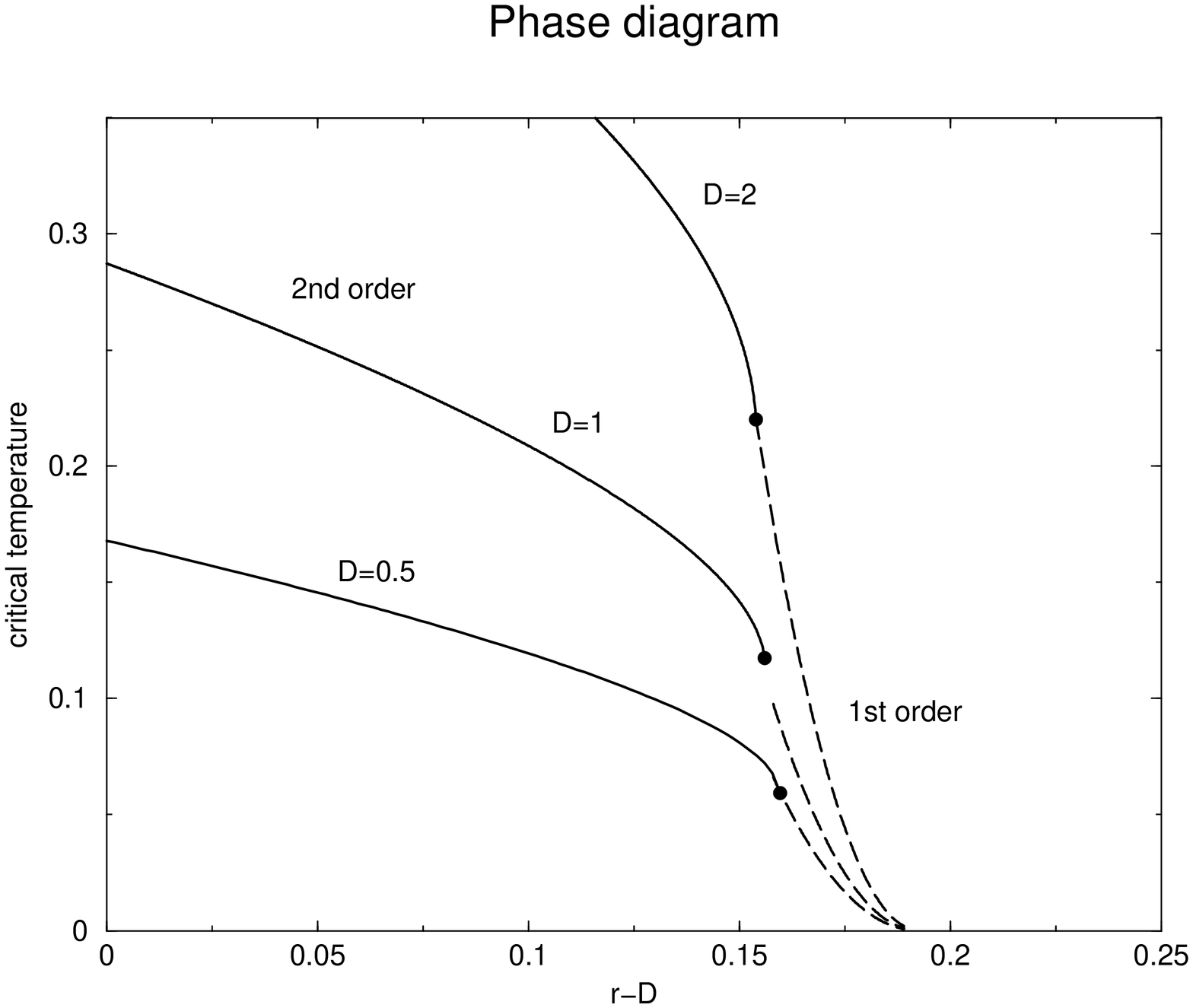}} &
\includegraphics[width=4cmi,angle=-90]{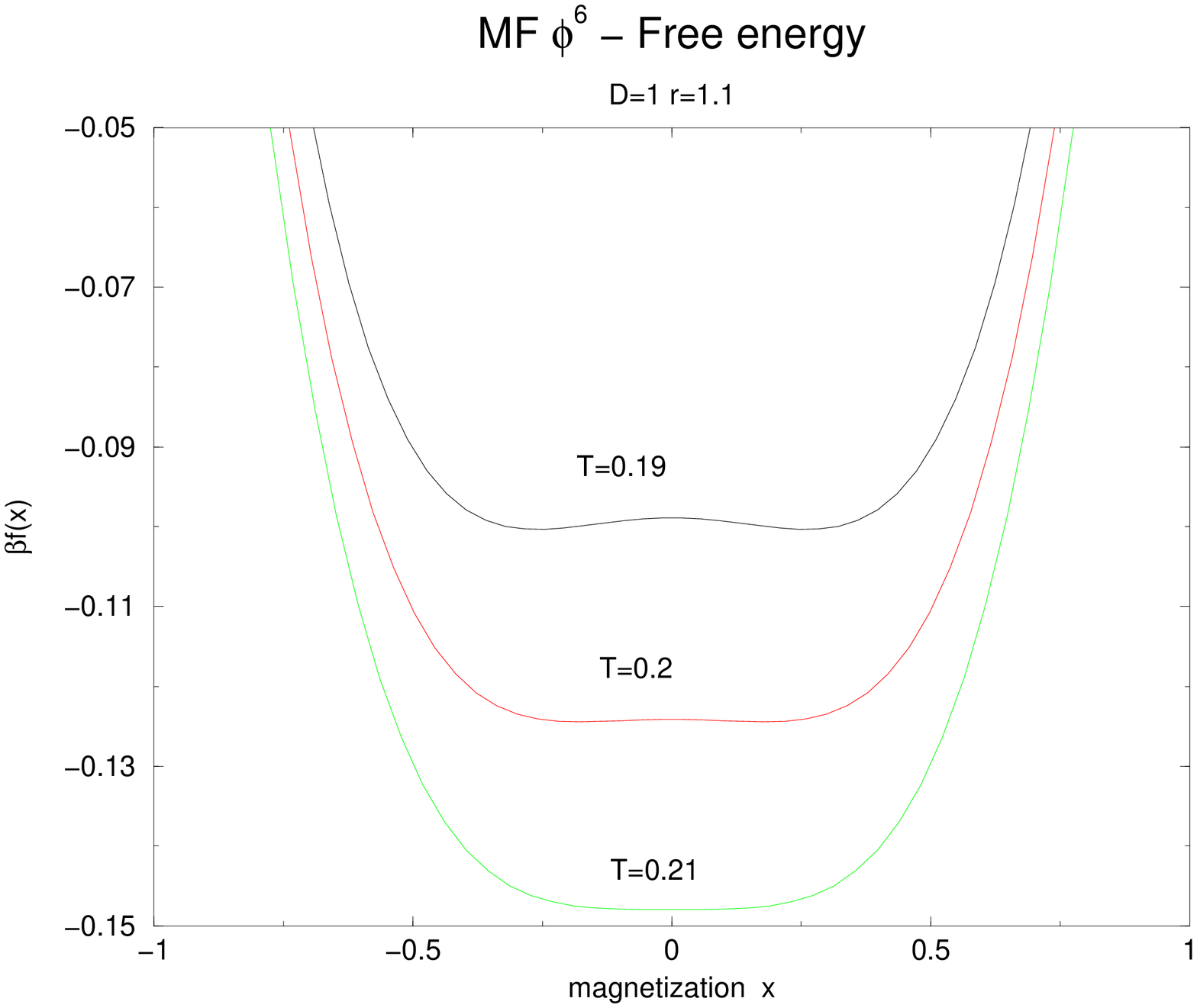}\\
\end{tabular}
\end{center}
\caption{Panel (a) shows the phase diagram of the $\phi^6$ model for
  different values of the coupling constant $D$. The solid
  (respectively dashed) line marks the second (resp.  first) order
  critical line, and the full dots the tricritical points. Panels (b)
  present the Landau free energy at a first order
  transition close to the tricritical point $T_{tr}$ and at a second
  order transition occurring on the line $a=0$, $T_c=0.205$. }
\label{phdiag}
\end{figure}

It is useful to consider the expansion up to sixth order of the Landau
free energy of the $\phi^6$-model in order to determine the critical line
of second order transitions and the tricritical point. We obtain
\begin{equation}
\beta f_L(x,\beta) = \frac{x^2}{2\beta D} - \psi(x,\beta) =
const. + \frac{ax^2}{2} +\frac{b x^4}{4} + \frac{c x^6}{6}+{\cal O}(x^8)~,
\end{equation}
where
\begin{eqnarray}
&& a(\beta,r,D) = \frac{1}{\beta D} -  \langle q^2 \rangle~, \\
&& b(\beta,r,D) = -\frac{\left(\langle q^4\rangle - 3 \langle q^2\rangle^2\right)}{6}~,
\end{eqnarray}
with
\begin{equation}
\langle q^m\rangle = {\int q^m \exp(-\beta V(q)) dq
 \over \int \exp(-\beta V(q)) dq}\quad.
\end{equation}
The numerical solution of $a=0$ yields the critical line of
second order transitions, see Fig.~\ref{phdiag}.  The tricritical
point, separating first and second order phase transition, is
determined by the more restrictive condition $a=b=0$.
Table~\ref{tableun} presents some values of the tricritical point
as a function of the parameter $D$.

\begin{figure}
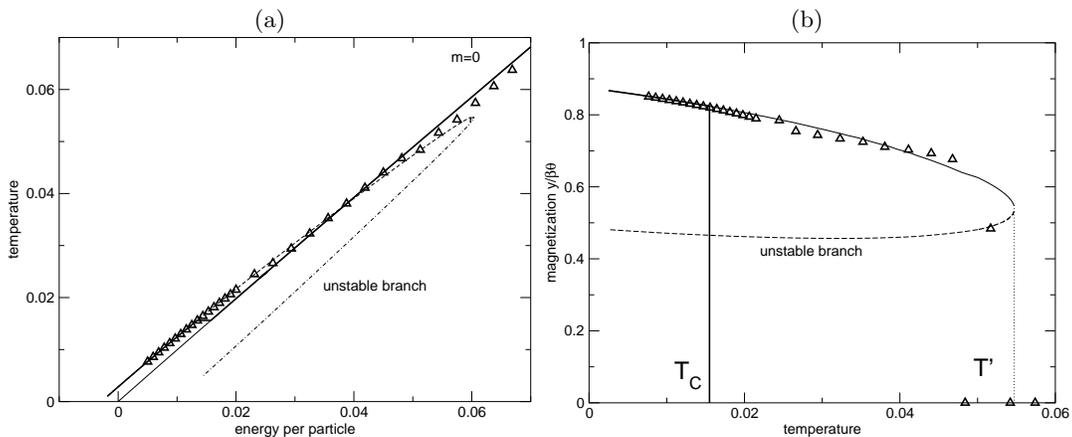

\begin{tabular}{cc}
(a) & (b)\\
\includegraphics[clip,width=70mm]
{phi6-tem-N1024.eps} &
\includegraphics[clip,width=70mm]
{phi6-mag-N1024.eps} \\
\end{tabular}
\caption{Thermodynamics of the $\phi^6$ model in the region of the
first-order transition ($r=1.18$, $D=1.0$): Panel (a) presents the caloric curve and
panel (b) the magnetization as a function of the energy. The
critical temperature is $T_c=0.0156 < T_{tr}$. Data obtained with
microcanonical simulations, $N=1024$ $t=1.25\ 10^6$.}
\label{p6mic}
\end{figure}

The canonical thermodynamics in the case of a first-order
transition is further illustrated in Fig.~\ref{p6mic}. Three branches of
solutions (two stable and one unstable) exist from $T=0$ up to
$T=T'$ where a saddle-node bifurcation occurs. The $m=0$ branch is stable
at all temperatures, while the stable (upper branch in Fig.~\ref{p6mic}) 
and unstable (lower branch in Fig.~\ref{p6mic})
$m \neq 0$ solutions meet and collide at $T=T'$. Notice that this is
at variance with the Blume-Emery-Griffiths model~\cite{beg} (BEG), where
the three branches do not extend down to zero temperature.

We have performed some simulations in the canonical ensemble to check this
caloric curve. The results are in agreement with the theory and, as
expected, display a marked metastability around the transition point
(hysteretic effects). More specifically, three different initial
conditions were adopted: (i) all $q_i=0$ (ii) all
$q_i=x_+$ (iii) random distribution between $q_i=0$ and
$q_i=x_+$. In all cases, the $p_i$ were initially chosen according to
a random gaussian distribution. Some microcanonical data are reported in
Fig.~\ref{p6mic} for an initial condition of type (ii).

A peculiarity of this model appears in some region of the parameters,
when one considers the caloric curves. Indeed one notes in
Fig.~\ref{p6mic}a that the $m=0$ line (full) crosses the magnetized curve
(dashed) to the left of $T'$, the temperature corresponding to the saddle node
bifurcation shown in Fig.~\ref{p6mic}b. This leads to the
impossibility of applying the usual Maxwell construction. However, this
is not always the case and, for example, $D=1$, $r=1.157>r_{tr}$ leads
to the usual features of a crossing to the right of $T'$. This
confirms that in the interval $r\in[r_{tr},1.157]$, the mean field $\phi^6$
model has a narrow region of negative specific heat, where the transition will be
first order in the canonical ensemble and second order in the
microcanonical. Unfortunately, the points near $T'$ are
extremely difficult to obtain because of numerical inaccuracies, and we
are therefore unable to report a clear determination of negative specific
heat.
In conclusion, this model shows a scenario similar to the
BEG model~\cite{beg}, in the case of a {\em dynamical} Hamiltonian.
Meanwhile, similar results have been published for some extensions of
the HMF model~\cite{antorc}.




\section{Conclusion}

The Blume-Emery-Griffiths mean-field model was shown to be
an excellent benchmark to discuss relations between canonical and
microcanonical ensembles in long range interacting systems~\cite{beg}.  Indeed,
this model is exactly solvable in both ensembles and is, at the same
time, sufficiently rich to display such interesting features as
negative specific heat and temperature jumps in the microcanonical
ensemble.  However, it has no dynamics and only the thermodynamical
behavior can be investigated. This is why we
need to study models that displays all these interesting
thermodynamical effects, but for which one would also dispose of an
Hamiltonian dynamics.  This point was already addressed in the framework
of the HMF model and in particular in its two-dimensional version
(see~\cite{hmfspringer} for a review).  Having access to dynamics, one
can moreover study non equilibrium features.

The mean-field models that we have considered in this paper are exactly solvable
in the canonical ensemble by a Hubbard-Stratonovich
transformation. The data in the microcanonical ensemble were
obtained using molecular dynamics simulations. We have shown that
these models have first and second order phase transitions and
tricritical points. Their phase diagram allows to test the presence
of ensemble inequivalence near canonical first order phase
transitions and we have also studied spontaneously generated 
out-of-equilibrium structures.

\section*{Acknowledgement} We would like to warmly thank Arkady
Pikovsky and David Mukamel for the helpful discussions which are at the
origin of this study.  This work has been partially supported by EU
contract No.  HPRN-CT-1999-00163 (LOCNET network), the P{\^o}le
Scientifique de Mod{\'e}lisation Num{\'e}rique de l'{\'E}NS Lyon, and the R{\'e}gion
Rh{\^o}ne-Alpes through the {\it Bourse d'Acqueil} nr. 00815559.
This work is also part of the
contract COFIN00 on {\it Chaos and localization in classical and
quantum mechanics}.

\begin{table}
\begin{tabular}{ccc}
$D$ & $r-D$   &  $T_{tr}$\\
\hline
0.5 &0.15965  & 0.05926 \\
1   & 0.156068 &  0.115159\\
2   &0.15393  &0.22018\\
5   &0.15247  &0.52811\\
\end{tabular}
\caption{Some numerical values of the tricritical points of the
$\phi^6$ model. Notice that $T_{tr}$ is approximatively proportional
to the coupling constant $D$.}
\label{tableun}
\end{table}

\begin{thebibliography}{99}

\bibitem{padmanabhan} T.~Padmanabhan, Phys. Rep., {\bf 188},
285 (1990), {\em Statistical mechanics of gravitating systems}.

\bibitem{springer} T. Dauxois, S. Ruffo, E. Arimondo, M. Wilkens
  (Eds), Lecture Notes in Physics 602, Springer (2002) {\em Dynamics and
  Thermodynamics of Systems with Long Range Interactions}.

\bibitem{Largedev} A. Dembo and O. Zeitouni {\em Large deviation techniques
and their applications}, Springer-Verlag (1998).

\bibitem{largedeviations} J. Barr{\'e}, F. Bouchet, T. Dauxois, S. Ruffo,
in preparation (2003).

\bibitem{Spohn} H. Spohn, {\em Large scale dynamics of interacting particles},
Springer-Verlag (1991).

\bibitem{hmfspringer}
T.~Dauxois, V. Latora, A. Rapisarda, S. Ruffo,
A. Torcini,  Lecture Notes in Physics 602 (Springer),
{\em The Hamiltonian mean field model: from dynamics to
statistical mechanics and back}.

\bibitem{zwanzig}
R.C. Desai, R. Zwanzig, J. Stat. Phys. {\bf 19}, 1 (1978),
{\em Statistical Mechanics of a nonlinear stochastic model}.

\bibitem{waterbag}
J. D. Dawson, Phys. Fluid {\bf 5}, 445 (1962), {\em One-dimensional
  plasma model};
Rev. Mod. Phys. {\bf 55}, 403 (1983), {\em Particle simulation of plasma}.

\bibitem{yoshida} H. Yoshida, Phys. Lett. A {\bf 150}, 262 (1990),
{\em Constructions of higher order symplectic integrators}.


\bibitem{barre} J. Barr{\'e}, F. Bouchet, T. Dauxois, S. Ruffo,
Phys. Rev. Lett. {\bf 89}, 110601 (2002). {\em
Out-of-equilibrium states as statistical equilibria of an
effective dynamics}.

\bibitem{barreepjb} J. Barr{\'e}, F. Bouchet, T. Dauxois, S. Ruffo,
European Physical Journal B {\bf 29}, 577 (2002).
{\em  Birth and long-time stabilization of out-of-equilibrium
coherent structures.}


\bibitem{lawden}
D. F. Lawden, {\em Elliptic Functions and Applications},
Springer (1989).

\bibitem{whittaker}
E. T. Whittaker, G. N. Watson,
{\em A Course of Modern analysis},
4th edition, Cambridge University Press (1946).

\bibitem{Arscott}
F. M. Arscott, {\em Periodic Differential Equation}, Pergamon Oxford (1964).

\bibitem{pecelli}
G. Pecelli, E. D. Thomas, Quarterly of Applied Mathematics {\bf 36},
129 (1978), {\em An example of elliptic stability with large
  parameters Lam{\'e}'s equation and the Arnold-Moser-R\''ussmann criterion}.


\bibitem{antoni} M. Antoni, S. Ruffo, Phys. Rev. E {\bf 52} 2361
  (1995), {\em Clustering and relaxation in Hamiltonian long-range
    dynamics}.

\bibitem{dauxois} T. Dauxois, P. Holdsworth, S. Ruffo, Eur. Phys. J.
  B, {\bf 16}, 659 (2000), {\em Violation of ensemble equivalence in
    the antiferromagnetic mean-field XY model.}

\bibitem{macla} R.I. MacLachlan and P. Atela, Nonlinearity {\bf 5}, 541
(1992), {\em The accuracy of symplectic integrators}.

\bibitem{Nose} S. Nos{\'e}, J. Chem. Phys. {\bf 81} 511 (1984), {\em
    Unified formulation of the constant temperature molecular-dynamics
    methods}; W.G.  Hoover, Phys. Rev. A {\bf 31} 1695 (1985), {\em
    Canonical dynamics- Equilibrium phase-space distributions}.

\bibitem{Tsallis} V. Latora, A. Rapisarda and C. Tsallis, Phys. Rev. E {\bf 64}
056134 (2001), {\em Non-Gaussian equilibrium in a long-range Hamiltonian system}.

\bibitem{beg} J. Barr{\'e}, D. Mukamel, S. Ruffo, Phys. Rev. Lett. {\bf
    87} 030601 (2001), {\em Inequivalence of ensembles in a system
    with long-range interactions}.

\bibitem{antorc} M. Antoni, S. Ruffo and A. Torcini, Phys. Rev. E
Rapid Communications {\bf 66}, 025103 (2002)
{\em First and second order clustering transitions for a system with
infinite-range attractive interactions}


\end{thebibliography}
\end{document}